\documentclass[apj]{emulateapj}

\def\Journal#1#2#3#4{{#4}, {#1}, {#2}, #3}

\def\NAT{Nature}
\def\AAA{A\&A}
\def\ApJ{ApJ}
\def\AJ{Astronom. Journal}
\def\Aph{Astropart. Phys.}
\def\ApJS{ApJSS}
\def\ML{Machine Learning}
\def\MNRAS{MNRAS}

\def\NIMA{Nucl. Instrum. Methods A}

\def\HESS{\mbox{H.E.S.S. }}

\begin{document}
\title{Observations of Mkn~421 with the MAGIC Telescope}

%
\author{
 J.~Albert\altaffilmark{a}, 
 E.~Aliu\altaffilmark{b}, 
 H.~Anderhub\altaffilmark{c}, 
 P.~Antoranz\altaffilmark{d}, 
 A.~Armada\altaffilmark{b}, 
 M.~Asensio\altaffilmark{d}, 
 C.~Baixeras\altaffilmark{e}, 
 J.~A.~Barrio\altaffilmark{d},
 H.~Bartko\altaffilmark{g}, 
 D.~Bastieri\altaffilmark{h}
% ...
}
 \altaffiltext{a}{Universit\"at W\"urzburg, D-97074 W\"urzburg, Germany}
 \altaffiltext{b}{Institut de F\'\i sica d'Altes Energies, Edifici Cn., E-08193 Bellaterra (Barcelona), Spain}
 \altaffiltext{c}{ETH Zurich, CH-8093 Switzerland}
 \altaffiltext{d}{Universidad Complutense, E-28040 Madrid, Spain}
 \altaffiltext{e}{Universitat Aut\`onoma de Barcelona, E-08193 Bellaterra, Spain}
% ...
\altaffiltext{g}{Max-Planck-Institut f\"ur Physik, D-80805 M\"unchen, Germany}
 \altaffiltext{h}{Universit\`a di Padova and INFN, I-35131 Padova, Italy} 
% ...


\begin{abstract}

The MAGIC telescope took data of very high energy $\gamma$-ray emission from the
blazar Markarian~421 (Mkn~421) between November 2004 and April 2005.  We present a combined
analysis of data samples recorded under different observational conditions,
down to $\gamma$-ray energies of  100~GeV. The flux was found to vary between
0.5 -- 2~Crab units (integrated above 200 GeV), considered a low state
when compared to known data.  Although the flux varied on a 
day-by-day basis, no short-term variability was observed, although there is
some indication that not all nights are in an equally quiescent state.
The results at higher energies
were found to be consistent with previous observations.  
A clear correlation is
observed between $\gamma$-rays and X-rays fluxes,  whereas no significant
correlation between $\gamma$-rays and optical data is seen. The spectral energy
distribution between 100~GeV and 3~TeV shows a clear deviation from a power
law, more clearly and at lower flux than previous observations at higher energies. 
The deviation persists after
correcting for the effect of attenuation by the extragalactic background
light, and most likely is source--inherent.  There is a rather clear indication of
an inverse Compton peak around 100~GeV. 
The spectral energy distribution of Mkn~421 can be fitted by a one-zone synchrotron self-compton 
model suggesting once again a leptonic origin of the very high energy $\gamma$-ray emission from this
blazar.

\end{abstract}


\section{Introduction}

Mkn~421 (redshift $z$ = 0.030) is the closest known and, along with Mkn~501,
the best studied TeV $\gamma$-ray emitting blazar.  
It was the first extragalactic 
source detected in the TeV energy range using imaging atmospheric Cherenkov telescopes (IACTs)
\citep{punch,petry}.  Mkn~421 is currently the source with the fastest observed flux
variations among TeV $\gamma$-ray emitters. So far it has shown flux variations larger 
than one order of magnitude, and occasional flux doubling times as short as
15~min \citep{gaidos,hegra421}.  Variations in the hardness of the TeV
$\gamma$-ray spectrum during flares were reported by several groups 
(e.g. \citet{whipple421,hess421}).
Simultaneous observations in the X-ray and~GeV-TeV bands show strong evidence for
flux correlation \citep{krawczynski,blazejwski}.  

Mkn~421 has been detected and studied in all accessible wavelengths of the
electromagnetic spectrum from radio waves to very high energy (VHE) $\gamma$-rays.  The
overall spectral energy distribution (SED) shows a typical two bump structure
with the first peak 
in the keV energy range and the second maximum at~GeV-TeV energies.
The SED is commonly interpreted as beamed, non-thermal
emission of synchrotron and inverse-Compton radiation from ultrarelativistic
electrons, accelerated by shocks moving along the jets at relativistic bulk speed.
Simple one-zone synchrotron-self-Compton (SSC) models (e.g.
\citet{coppi,costamante}) describe the observational results
satisfactorily well.  However, hadronic models \citep{mannheim,muecke} can
explain the observed features, too.  A way to
distinguish between the different emission models is to determine the position of 
the second peak in the SED,
using simultaneous time-resolved data over a broad energy range through 
multiwavelegth campaigns. This requires providing data in the as yet unexplored
gap in the SED.

The MAGIC telescope ({\bf M}ajor {\bf A}tmospheric {\bf G}amma {\bf I}maging
{\bf C}herenkov telescope; see \citet{magic}), located  on the Canary Island
La~Palma (2200 m asl, 28$^\circ$45$'$N, 17$^\circ$54$'$W), completed its
commissioning phase in early fall 2004. 
MAGIC is currently the largest IACT,
with a 17~m diameter tessellated reflector dish consisting of 964
0.5~$\times$~0.5~m$^2$ diamond-milled aluminium mirrors. Together with the
current configuration of the MAGIC camera with the trigger region of 2.0 degrees 
diameter \citep{cortina}, this results in a trigger collection area for
$\gamma$-rays of the order of $10^5$~m$^2$, increasing with the zenith
angle of the observation. Presently the accessible trigger energy range spans from
50-60~GeV (at small zenith angles) up to tens of TeV.  The MAGIC telescope is
focused to 10 km distance 
-- the most likely position for a 50~GeV air shower.
The accuracy in reconstructing the direction of incoming
$\gamma$-rays on an event by event basis, 
hereafter $\gamma$ point spread function (PSF), is about
$0.1$~degrees, slightly depending on the analysis.  

The first physics observations in winter 2004/05 and in spring 2005 included
observations of the well established TeV blazar Mkn~421. In total, 19 nights of
data were taken on this source, the observation times per night ranging from 30
minutes up to 4 hours. 
Here we present the results from these observations, covering the 
energy range from 100~GeV to several TeV.
We first describe the data set and analysis techniques
in section~\ref{observations}. In section~\ref{results}, we present the
results and, finally, in section~\ref{discus}, we compare our results with other
observations and interpret them in terms
of different models.

\section{Observations and data analysis}\label{observations}

\begin{table*}
\begin{center}
\caption{\label{tab:data} Results of the Mkn~421 data using the $Alpha$ approach 
(see text for details). 
Samples I+II were recorded in November 2004 - January 2005, while samples III+IV were taken in 
April 2005.}

\vspace{1mm}

\begin{tabular}{ c c c c c c c c c }
\hline
{\bf \scriptsize {sample}} & {\bf \scriptsize on time} & 
{\bf \scriptsize ZA range \hspace{1mm} [$^{\circ}$]}  &
{\bf \scriptsize mode} &
{\bf \scriptsize $E_{thr}$[GeV]} & {\bf \scriptsize N$_{\mathrm{on}}$} & 
{\bf \scriptsize N$_{\mathrm {off}}$} & 
{\bf \scriptsize N$_{\mathrm {excess}}$ } & {\bf \scriptsize sigma} \\
\hline 
\hline
\normalsize
 I   & 4.63 h & 9.3 - 31.2  & ON     & 150 & 3761 & 1878 $\pm$ 32 & 1883 $\pm$ 69 & 29.3 \\
\hline
 II  & 1.53 h & 42.4 - 55.0 & ON     & 260 & 1086 &  674 $\pm$ 25 &  413 $\pm$  41 & 10.1 \\
\hline
 III & 9.30 h & 9.2 - 27.5  & ON     & 150 & 8083 & 4360 $\pm$ 49 & 3723 $\pm$ 102 & 38.9 \\
\hline
 IV  & 10.12 h & 9.4 - 32.4  & wobble & 150 & 7740 & 4532 $\pm$ 67 & 3208 $\pm$  111 & 29.1 \\
\hline
\end{tabular}
\end{center}
\end{table*}

The Mkn~421 data were taken between November 2004 and April 2005, and divided into
four samples, for reasons given below. Data taken before and after 
February 2005 were treated separately, 
due to changes in the telescope hardware.  
Most of the data were taken at small zenith angles ($ZA < 30^\circ$), 
i.e. at a low trigger energy
threshold.  However, observations made during 1.5~hours in a common campaign 
with the \HESS telescope system~\citep{hess_magic} in December 2004
were taken at $42^\circ < ZA < 55^\circ$.
There were also different observational modes: the standard mode for MAGIC is the
ON-OFF mode, with equal time given to tracking the source in
the center of the camera (ON), and tracking a sky region near the source but 
with the source outside the field of view (OFF).
This provides a robust estimate of the background.
In our observations, we considered the $\gamma$-ray signal from Mkn~421 to be
strong enough to obviate OFF observations, and we estimated the background
level from the ON data (see below).  In April 2005, part of
the data were taken in the wobble mode \citep{daum}.  In this mode, two sky
directions, opposite and 0.4$^\circ$ off the source, were tracked alternately
for 20 minutes each, which provides a simultaneous measurement of signal and
background. In the wobble mode there is {\em a~priori} no need for additional OFF data.

\begin{figure}[h]
\begin{center}
\begin{minipage}[b]{0.45\textwidth}
\includegraphics*[width=1\textwidth,angle=0,clip]{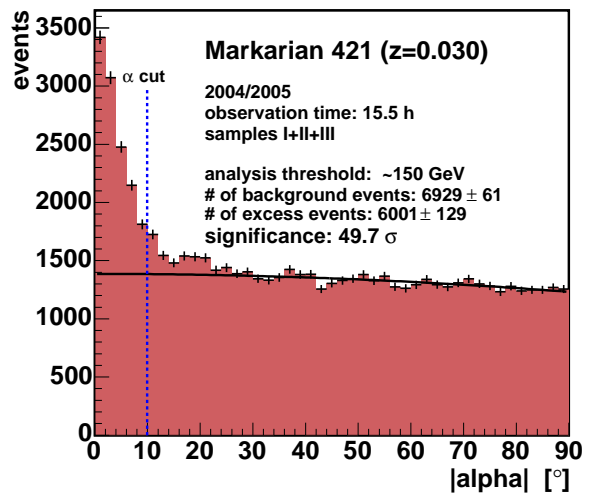}
\caption{\label{fig_alpha} $Alpha$ distribution for the combined data
	samples I+II+III with $E_{thresh}=150$~GeV. A vertical line indicates the 
    $Alpha$ cut used to extract excess events. The black parabola is a fit 
    to the $Alpha$ distribution between 30 and 90 degrees and is 
    used to estimate the background level between 0 and 10 degrees.}
\end{minipage}
\end{center}
\end{figure}

\begin{figure}
\begin{center}
\begin{minipage}[b]{0.40\textwidth}
\includegraphics*[width=1\textwidth,angle=0,clip]{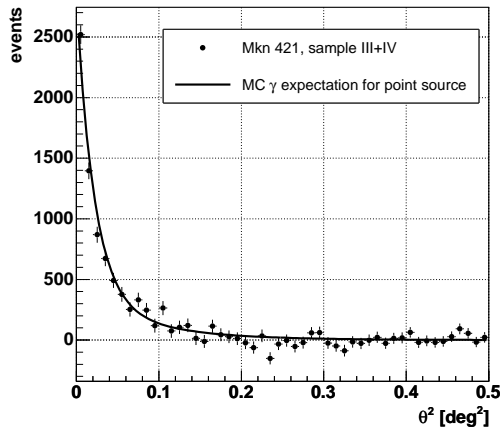}
\caption{\label{fig_theta2} $\theta^{2}$ distribution for the combined data
	samples III+IV with $E_{thresh}=150$~GeV after background subtraction. 
    The black line is the MC-$\gamma$ expectation for a point source.}
\end{minipage}
\end{center}
\end{figure}

\begin{figure}
\begin{center}
\begin{minipage}[b]{0.45\textwidth}
\includegraphics*[width=1\textwidth,angle=0,clip]{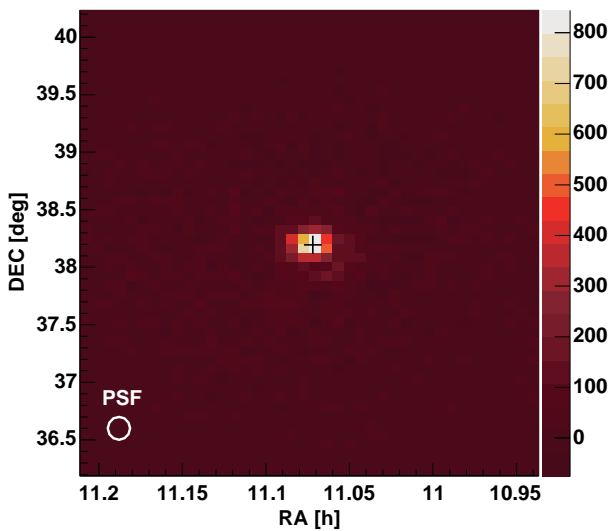}
\caption{\label{fig_skymap}
         Sky map of excess events in the region of Mkn~421 
         for samples III+IV using the $Disp$ method~\citep{domingo}.
         The black cross indicates the reconstructed source position.
         Note that the vertical scale is in units of [events / (0.05$\times$0.05 deg$^2$)].}
\end{minipage}
\end{center}
\end{figure}

The observation criteria and some important parameters of the four 
data samples are summarized in Table~\ref{tab:data}.  For
each data sample a separate Monte-Carlo (MC) set of $\gamma$ events was
simulated (CORSIKA version 6.023, \citet{corsika,majumdar}), taking into
account the zenith angle of observation, the observational mode, and the
hardware setup of the telescope.
The full data set corresponds to 29.0~hours ON-source observation time. 
Runs with problems in the hardware or
with unusual trigger rates were rejected in order to ensure a stable performance and
good atmospheric conditions. After removing these runs, the remaining observation
time was 25.6~h.

For calibration, image cleaning, cut optimization, and energy reconstruction
the standard analysis techniques of the MAGIC telescope \citep{bretz,rwagner,gaug} 
were applied as shortly described below.  
The calibration of the raw data from
the MAGIC camera uses a  system  consisting of fast
and powerful LED pulsers emitting at three different wavelengths with variable 
light intensity.
Absolute calibration is obtained by comparing the signal of the pixels with 
the one obtained from a carefully calibrated
PIN diode, and is cross-checked by analysing muon rings.  
The time resolution of the read-out system has been measured to be 
about 700~ps for Cherenkov light flashes of 10 photo-electrons (ph.el.) per pixel,
reaching 200~ps at 100 ph.el.  Calibration
events are taken at 50 Hz, interlaced with normal data, 
using an external calibration trigger.

The calibrated images are cleaned using so-called tail cuts: pixels are retained only 
if their reconstructed charge signals are larger
than 10~ph.el. ('core pixels') or if their charges are larger
than 5~ph. el. and they have at least one neighboring core pixel. 
The camera images are then reduced to image parameters as in~\citep{hillas}, adding
parameters describing the intensity concentration and asymmetry.

\begin{figure}[h]
\begin{center}
\begin{minipage}[b]{0.45\textwidth}
\includegraphics*[width=1\textwidth,angle=0,clip]{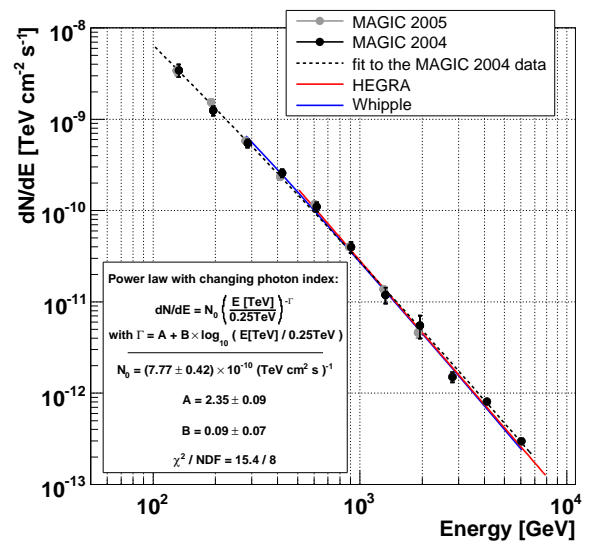}
\caption{\label{fig:crab} Differential energy distribution of the Crab Nebula data sample
    from 2004 (black circles) and 2005 (grey circles) as measured by MAGIC \citep{rwagner}.  A fit by power law with a changing 
    photon index 
    to the MAGIC 2004 data is shown by the dashed black line. The analytical form of the fit
    and the fit parameters are listed in the inlay.
    Wipple data (solid blue line, \citet{whippleCrab}) and HEGRA data (solid red line, \citet{hegraCrab}) are shown for comparison.}
\end{minipage}
\end{center}
\end{figure}

For $\gamma$/hadron separation a multidimensional
classification technique based on the Random Forest (RF) method \citep{breiman,rf}
was used.  The RF method uses training data (randomly chosen data events and Monte Carlo $\gamma$s,
representing background and signal)
to find a set of classification trees in the space of image parameters. Multiple
trees are combined to form a generalized predictor by taking the mean classification
from all trees. The predictor, called hadronness, spans a range between 0 and 1, 
and characterizes the  event images being less or more hadron-like.

In our analysis, classical image shape parameters like $Width$, $Length$, $Dist$ and
$Size$ were used as input parameters. The cuts in hadronness 
for the $\gamma$/hadron
separation were trained for each data set separately, and 
were then chosen such that the overall cut efficiency
for MC $\gamma$ events remained about 50$\%$. The corresponding hadron suppression 
is about 90-99$\%$, improving  with increasing $Size$ of the events.

A critical variable not used in the RF classification tree is
$Alpha$, the angle between the major image axis and the line connecting the center of
gravity of the image with the source position in the camera plane.
In stand-alone IACTs, $Alpha$ is commonly used, 
after all previously noted cuts, to extract the $\gamma$ signal
from the data, and to estimate the level of background.  
For a point source, the $Alpha$ distribution of the
$\gamma$-like events is expected to peak at low
values of $Alpha$, whereas for background events the distribution should
be flat or slowly varying with Alpha.

In the case of our ON-mode data, the background remaining after $\gamma$/hadron
separation was estimated from the $Alpha$ distribution by performing a second
order polynomial fit (without linear term) 
in the range between 30$^\circ$ and 90$^\circ$ 
where no contribution from
$\gamma$ events is expected (see Fig.~\ref{fig_alpha}).  The signal was then
determined as the number of observed events in the range $Alpha < Alpha_0$
exceeding the fit extrapolated to small $Alpha$, where $Alpha_0$ is energy
dependent and has a typical value of 15$^\circ$.  
The significance of an excess is then calculated according to Eq.~17 in \citet{LiMa}.

In the wobble mode, the ON
({\em source}) data are defined by calculating image parameters with respect to
the source position, whereas OFF data are obtained from the same events but with image
parameters calculated with respect to the position on the opposite side of the
camera, the {\em antisource} position.  In order to avoid an unwanted
contribution from $\gamma$-events in the OFF sample and to guarantee the statistical
independence between the ON and the OFF samples in the signal region, 
the following procedure is applied:
events with $Alpha_{source} < Alpha_0$ (with $Alpha_{source}$ calculated with respect to the
source position) are excluded from the OFF sample, and events with $Alpha_{antisource} <
Alpha_0$ (with $Alpha_{antisource}$ calculated with respect to the antisource position)
are excluded from the ON sample.
This cut assures that the $Alpha$ distributions for ON
and OFF events are statistically independent for $Alpha < Alpha_0$.  
The $Alpha$ approach was used to determine the excess events for all four data sets
(Table~\ref{tab:data}). 

As an alternative to this classical approach using $Alpha$, the so-called
$\theta^{2}$ approach can be applied, an approach more common for the analysis of data from
a system of IACTs like HEGRA or \mbox{H.E.S.S}.  The angle $\theta$ denotes the angular
distance between the expected source position and the reconstructed origin of the
initial $\gamma$-ray.  Since for a single IACT the angle $\theta$ can not be reconstructed 
directly, the so-called $Disp$ method \citep{fomin,lessard,domingo} was used to determine
the source position in the camera plane, using position-independent image shape parameters. 
The number of excess events is then determined as
the difference between the $\theta^{2}$
distributions for the source and background, respectively,
similar to the $Alpha$ approach. 
The background-subtracted $\theta^{2}$ distribution for samples III and IV
is shown in Fig.~\ref{fig_theta2}. The average background was estimated from
the wobble data themselves, by excluding the sector of the camera affected by the 
presence of the strong source. The solid line in Fig.~\ref{fig_theta2} indicates the expectation
from MC-$\gamma$ events for a point source.  
Computing $\theta^{2}$ also permits
to produce sky maps in which for every 
$\gamma$-ray candidate an origin in the sky is assigned (see Fig.~\ref{fig_skymap}).
Note that our signal analysis relies on the $Alpha$ approach throughout.

These conservative analysis methods are known to produce reliable results for energies
above 100~GeV.  The energy regime below 100~GeV 
will require additional studies, in particular concerning the background
rejection.  
Thus, for our analyzed sample the $Size$ parameter (total amount of light of the image and
in first order proportional to the energy) was required to be
above 150 photoelectrons.
 
The energy estimation was performed using again the Random Forest
technique, based on the image parameters of a MC $\gamma$
sample. This sample is statistically independent of the one used for the
training of the $\gamma$/hadron separation cuts. Prior to the training of
the energy estimation, loose (high-acceptance) cuts in hadronness and $Alpha$ 
were applied to avoid a possible bias caused by outlier $\gamma$ events.

The energy thresholds of the individual analyses (as given in
Table~\ref{tab:data}) are defined as the peak in the differential energy
distribution of the MC-$\gamma$ events after all cuts.  
Our analyses showed that we were able to extract excess events with energies
$\sim$50~GeV lower than the corresponding peak value.

The spectrum of the number of excess events in bins of true energy is determined
from the spectrum in the estimated energy by an unfolding procedure. This
procedure corrects for the finite energy resolution and for biases in the
energy estimation. The unfolding program package used in MAGIC allows 
unfolding with a variety of methods (\citet{anykeyev}), which differ in the way
regularization is implemented. Unfolding results are only accepted if the
results from the different methods are consistent with each other and if some
criteria are satisfied concerning the regularization strength, the size of
the noise component and the $\chi^2$ value. The latter is a measure of the agreement
between the expected ``measured'' spectrum from the unfolded spectrum and the
actually measured spectrum. The unfolding result presented in this analysis was
obtained with an iterative method, as described in \citet{bertero}.

To demostrate the quality of the applied analysis and the good agreement with
previous measurements by other experiments, we show in Fig.~\ref{fig:crab} the
differential energy spectrum of the Crab Nebula data (``standard candle'' of
VHE $\gamma$-ray atronomy).  The data were taken in 2004 and 2005 with
observation conditions and telescope perfomance similar to those of the Mkn~421
data.  Additional publications describing details of the calibration methods
and the data analysis are in preparation.

\begin{figure*}
\centering
\includegraphics[bb=20 0 568 270,width=0.95\textwidth,angle=0,clip]{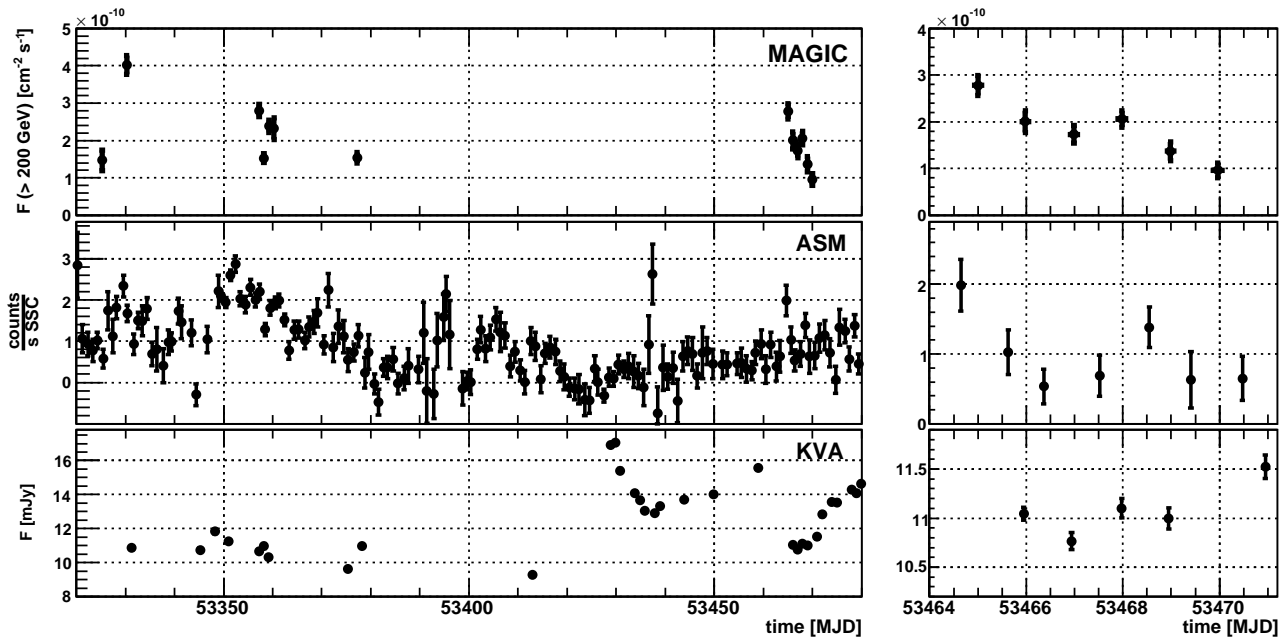}
\caption{\label{fig_lightcurve} Night-by-night light curve for Mkn~421 from November 2004
to April 2005. {\it Left panels}: data from November 2004 to April 2005.
{\it Right panels}: expanded data for 6 nights in April 2005. 
{\it Upper panel}: MAGIC data, night average of Mkn~421 above 200~GeV using samples I+III+IV.  
{\it Middle panel}: corresponding day-by-day X-ray counts as observed by the RXTE/ASM.  
{\it Lower panel}: Light curve of the optical flux of Mkn~421 as measured by the KVA telescope.}
\end{figure*}

\section{Results}\label{results}

\subsection{The signal}\label{results:signal}

During the entire observation period, Mkn~421 was found to be in a low 
flux state compared with existing data
(around 1~Crab unit for a flux integrated above 200 GeV), 
but resulting in a clear signal in all four data samples.
Fig.~\ref{fig_alpha} shows the $Alpha$ distribution of the $\gamma$-candidates
of the combined samples I, II, and III with an energy threshold of $\sim$150~GeV. An
excess of about 7000 events was found, which, for the given background, corresponds
to an excess of more than 49 standard deviations.
The number of excess events and the significances for the
individual samples are summarized in Table~\ref{tab:data}.

Fig.~\ref{fig_skymap} shows a sky map produced with the $Disp$
method using samples III and IV. 
The reconstructed source position from
the sky map (Fig.~\ref{fig_skymap}, indicated by the black cross)
is centered at RA=+11h04$'$19$''$, DEC=38$^\circ$11$'$41$''$. 
The systematic error of the telescope pointing
is 2\,$'$. The $\gamma$ PSF is indicated by a white circle in the left 
bottom corner. The observed extension
of Mkn~421 is compatible with the MC expectation of a point source, which can also 
be seen in Fig.~\ref{fig_theta2}.

\subsection{The light curve}\label{results:lc}

The integral fluxes above 200~GeV, averaged over each night of observation, are
shown in the upper panels of Fig.~\ref{fig_lightcurve}. Significant variations of
up to a factor of four overall and up to a factor two in between successive nights can
be seen.  Since sample II has an energy threshold of 260~GeV it is not shown on
the light curve.
The relatively high analysis energy threshold of 200~GeV applied for the light curve 
ensures that the results are independent of the actual trigger thresholds
during each night.  In the middle panels of Fig.~\ref{fig_lightcurve} the
corresponding flux in the X-ray band as observed 
by the All-Sky-Monitor (ASM\footnote{see http://heasarc.gsfc.nasa.gov/xte\_weather/})
on-board the RXTE satellite is shown.
In the lower panels of Fig.~\ref{fig_lightcurve} the
optical data taken by the KVA telescope\footnote{see http://tur3.tur.iac.es/}
on La Palma are shown.  Note that the
contribution of the host galaxy (appr. 8.0 mJy) has been subtracted. While the
X-ray data show a moderate variability within the observation period, the
optical flux stays almost constant.

\begin{figure*}
\begin{center}
\begin{minipage}[b]{0.95\textwidth}
\begin{center}
\includegraphics[width=.9\textwidth,angle=0,clip]{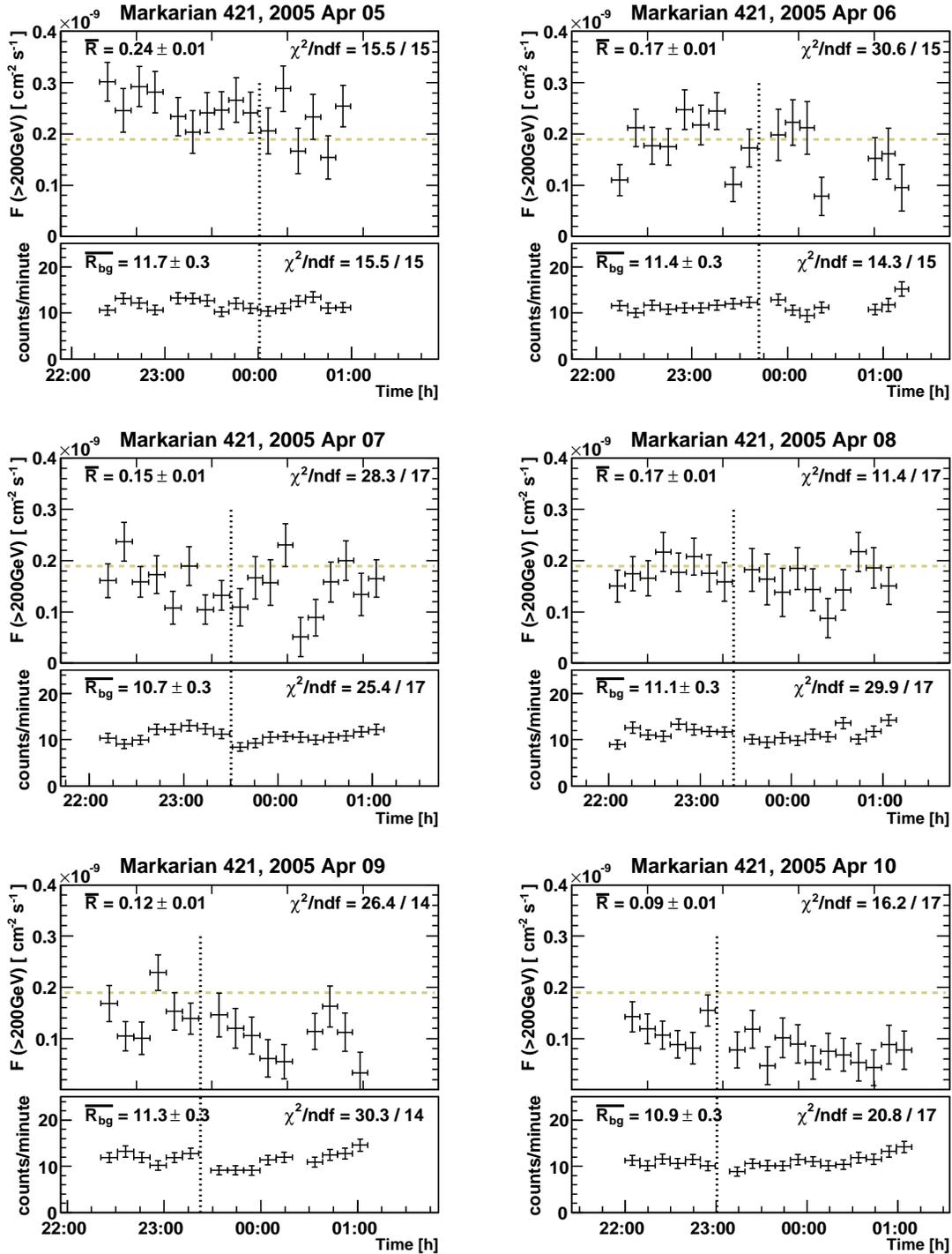}
\caption{\label{fig_lc6days}
         Light curve for 6 nights in 2006 April in 10 minutes binning. 
         {\it Upper panels}: flux above 200 GeV. Mean rate $\overline{\mbox{R}}$ in units
         of ($10^{-9} \mbox{cm}^{-2} \mbox{s}^{-1}$) and the quality of the fit by a 
         constant are shown in the panel.
         {\it Lower panels}: mean background rate  $\overline{\mbox{R}_{\mbox{bg}}}$ 
         per minute after cuts. Note the rising background rate towards the end of 
         each observation slot, related to the rising moon.
         $\overline{\mbox{R}_{\mbox{bg}}}$ and the quality of the fit by a 
         constant are shown.
         The vertical dotted lines indicate the time of the switchover from the ON 
         observational mode to the Wobble mode.
         The dotted horizontal line indicates the Crab Nebula integral flux above 
         200 GeV as measured by MAGIC \citep{rwagner}. }
\end{center}
\end{minipage}
\end{center}
\end{figure*}

For the 6 nights in April (MJD 53465 to 53471), the light curve above 200 GeV is
shown in Fig.~\ref{fig_lc6days} in bins of 10 minutes.  We also added the
background rates for each night in the same binning, in order to demonstrate
that the small variations in the excess rates and the daily changes are not
caused by detector effects or atmospheric transmission changes.  The vertical
lines indicate the time in each night at which the observation mode was changed
from ON to Wobble.  The mean integral flux per night $\overline{\mbox{R}}$ in
units of $10^{-9} \mbox{cm}^{-2} \mbox{s}^{-1}$ and the quality of the fit
constant are shown in the panels.  The horizontal dashed line corresponds to
the integral flux of the Crab Nebula above 200 GeV. Combining the findings from
the intra-night light curves we conclude that we did not find significant
short-term flux variability within individual nights, despite the high
sensitivity of MAGIC for such a search.  Some of the nights, however, are less
compatible with a constant flux than others, which might be an indication
of some activity, albeit unstructured and difficult to quantify.
On the other hand, we observe
significant day-to-day variation by up to a factor of two, and differences up
to a factor of four in the full sample.

\subsection{The energy spectrum}\label{results:spectrum}
\subsubsection{The measured spectrum}\label{results:measspectrum}

\begin{figure}
\begin{center}
\includegraphics*[width=0.45\textwidth,angle=0,clip]{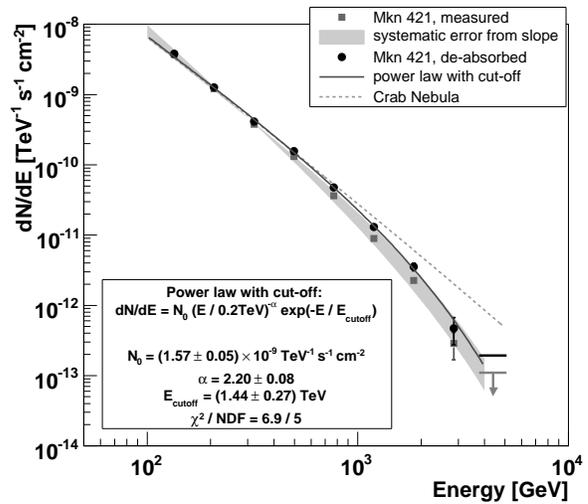}
\caption{\label{fig_spectrum}
         Differential energy distribution for Mkn~421 averaged over the whole data sample.
         The measured energy spectrum is shown by the grey full squares and the de-absorbed spectrum
         by the black full circles. The spectral point at the highest energy is a 95\% upper limit.
         The grey shaded area corresponds to a systematic error from a slope error of $\pm$0.2 as quoted in the text.
         The black solid line indicates the best fit to the de-absorbed 
         spectrum by a power law with exponential cut-off; its parameters are listed in the inlay.
         For comparison reasons, the measured Crab Nebula spectrum \citep{rwagner} is shown by 
         the grey dotted line.}
\end{center}
\end{figure}

For the spectrum calculation, we combined the entire data set because 
the differences between the fluxes on individual nights are rather moderate
(see Fig.~\ref{fig_lightcurve}).
The resulting averaged differential energy spectrum is shown in Table~\ref{tab:spec} 
and in Fig.~\ref{fig_spectrum} by filled grey boxes. The energy spectra extend from
around 100~GeV to several TeV. The last spectral point at 4.4~TeV is an
95\% upper limit. The error bars shown are statistical only.
Systematic errors are estimated to be 18\% on the absolute energy scale,
which correspond to 44\% on the absolute flux level for a photon index of 2.2.
The systematic error on the slope is estimated to be 0.2.
The attenuation of the VHE photons by intergalactic low energy photons and the 
determination of the intrinsic spectrum of Mkn~421 are discussed below.
 
\subsubsection{$\gamma$-ray absorption by the EBL}\label{results:ebl}

\begin{figure*}
\begin{center}
\begin{minipage}[b]{0.96\textwidth}
\includegraphics*[width=1.0\textwidth,angle=0,clip]{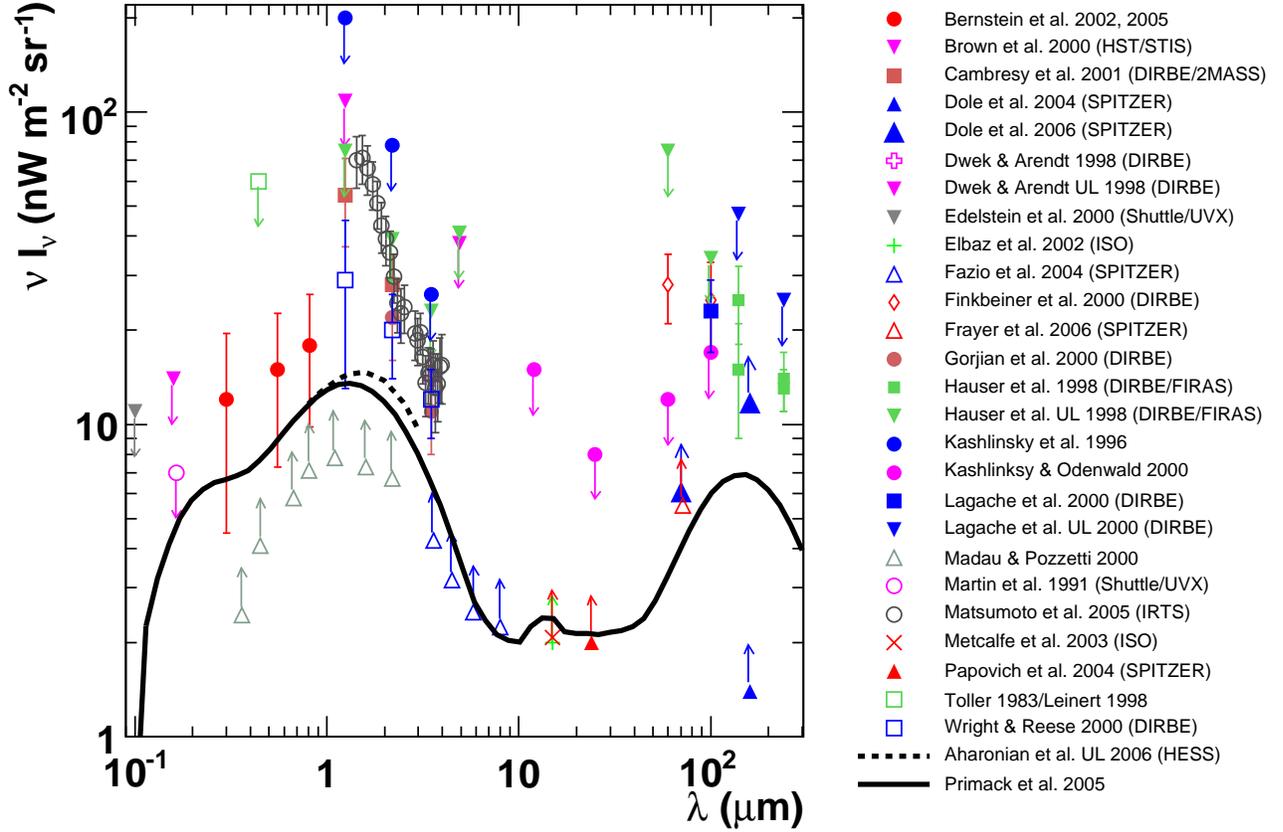}
\caption{\label{fig_ebl}
         Energy density of the extragalactic background light (EBL).
         Direct measurements, galaxy counts, low and upper limits are shown by 
         different symbols as described in the legend. The black 
         solid curve is the EBL spectrum as in \citet{primack} for z=0 
         but upscaled by a factor 1.5 to match low limits derived
         from the galaxy counts \citep{elbaz,metcalfe,spitzer}.} 
\end{minipage}
\end{center}
\end{figure*}

The VHE photons from Mkn~421 cross $\sim$400 million light years on their way
to Earth.  They interact with the low energy photons of the
extragalactic background light (EBL, see
\citet{nikishov,gould,stecker,HauserDwek}) consisting of redshifted 
star light of all epochs and reemission of a part of this light by dust in galaxies.
The most common reaction channel between VHE $\gamma$-rays and 
the low energy photons of the EBL is pair production
  $ \gamma_{\mbox{\scriptsize{\,VHE}}} + \gamma_{\mbox{\scriptsize EBL}} \rightarrow e^{+}\,e^{-}$,
a reaction which has its largest cross section
when the center of mass energy is roughly 3.6 times larger than the
threshold energy of 2m$_e$c.
The intrinsic (de-absorbed) photon spectrum, $dN/dE_i$, of a blazar located
at redshift $z$ is given by:
\[
  dN/dE_i\,=\, dN/dE_{obs} \times \exp[\tau_{\gamma\gamma}(E,\,z)],
\]
where $dN/dE_{obs}$ is the observed spectrum and 
$\tau_{\gamma\gamma}(E,\,z)$ is the optical depth.
The distance to Mkn~421 implies that the optical depth 
(e.g. Eq.~2 in \citet{dwek}) strongly depends on the shape 
and absolute photon density of the EBL between 1 and 30~$\mu$m wavelength.
A rather complicated distortion of the intrinsic spectrum takes
place above $\sim$100~GeV. 
Although the calculation of the optical depth is straightforward,
the spectral energy distribution of the EBL is uncertain. 
Direct measurements of the EBL are difficult because of the strong 
foreground emission consisting of reflected sunlight and thermal emission 
from zodiacal dust particles. 
Hence, many measurements lead to upper limits 
\citep{hauser98,dwekarendt}.
Several measurements claimed a direct detection of the EBL, but some of them 
are controversial \citep{matsumoto,finkbeiner}.
An alternative method to determine the EBL are fluctuation analyses of the measured 
radiation. Since a part of the EBL originates from discrete sources,
fluctuations in the number of sources in an observer's field of view
will produce fluctuations inthe measured background 
\citep{kashlinsky96,kashlinsky00}.
A third method is the galaxy number counting in the deep field surveys, 
which provides robust lower limits to the SED of the EBL
\citep{elbaz,metcalfe,spitzer,madau}. 
The results of these methods and measurements
are summarized in Fig.~\ref{fig_ebl}.

\begin{table}
\begin{center}
\caption{\label{tab:spec} Averaged measured differential energy spectrum of Mkn~421, derived
from the data set presented in the Paper. The last point is an 95\% upper limit.}

\vspace{3mm}

\begin{tabular}{ p{1.2cm} p{1.2cm} p{1.2cm} c }
\hline
 \multicolumn{3}{c}{energy bin [GeV] }  & differential flux dN/dE \\
lower bin limit & mean \mbox{energy} & upper bin limit & 
[\,$\mathrm{photons} / \left( \mathrm{TeV}\,\mathrm{cm}^2\,\mathrm{s} \right) $\,] \\
\hline 
\hline
108   & 134   &  167      &  $(3.72 \pm 0.34)\times10^{-9} $    \\
167   & 208   &  259      &  $(1.21 \pm 0.04)\times10^{-9} $    \\
259   & 321   &  402      &  $(3.77 \pm 0.15)\times10^{-10}$     \\
402   & 498   &  623      &  $(1.32 \pm 0.05)\times10^{-10}$     \\
623   & 770   &  965      &  $(3.63 \pm 0.19)\times10^{-11}$      \\
965   & 1192   &  1497    &  $(8.95 \pm 0.71)\times10^{-12}$     \\
1497   & 1845   &  2321   &  $(2.26 \pm 0.27)\times10^{-12}$      \\
2321   & 2856   &  3598   &  $(2.88 \pm 1.20)\times10^{-13}$     \\
3598   & 4429   &  5579 &  $ < 1.10 \times10^{-13}$                        \\
\hline
\end{tabular}
\end{center}
\end{table}

\begin{table}
\begin{center}
\caption{\label{tab:systcutoff} Systematic study of the fit parameters on the de-absorbed spectrum of Mkn~421.
The fitted fuction is a power law with an exponential cut-off:
$ dN/dE = N_{0} (E/0.2\,\mathrm{TeV})^{-\alpha} \exp(-E/E_{\mathrm{cutoff}}) $.
We show fit values on the photon index, $\alpha$, and the 
cut-off energy, $E_{\mathrm{cutoff}}$, for following assumptions:
nominal values (A), a systematic shift by +18\% in the VHE energy scale (B),
a systematic shift by -18\% in the VHE energy scale (C),
a systematic shift by +18\% in the VHE energy scale and, in addition, 25\% more density of the EBL (D),
and a systematic shift by -18\% in the VHE energy scale and, in addition, 25\% less density of the EBL (E).
Note that the resulting systematic errors are comparible with the statistical errors.
}

\vspace{5mm}

\begin{tabular}{  l c c }
\hline
                            &  $\alpha$ &  $E_{\mathrm{cutoff}}$ [TeV]  \\
\hline
  A: nominal                &  $2.20\pm0.08$  &  $1.44 \pm 0.28 $    \\
  B: E+18\%                 &  $2.16\pm0.08$  &  $1.59 \pm 0.29 $     \\
  C: E-18\%                 &  $2.24\pm0.08$  &  $1.26 \pm 0.26 $     \\
  D: (E+18\%) + (EBL+25\%) &  $2.12\pm0.08$  &  $1.61 \pm 0.29 $      \\
  E: (E-18\%) + (EBL-25\%) &  $2.20\pm0.08$  &  $1.09 \pm 0.20 $     \\
\hline
\end{tabular}
\end{center}
\end{table}

\begin{figure}
\begin{center}
\begin{minipage}[b]{0.46\textwidth}
\includegraphics*[width=1.0\textwidth,angle=0,clip]{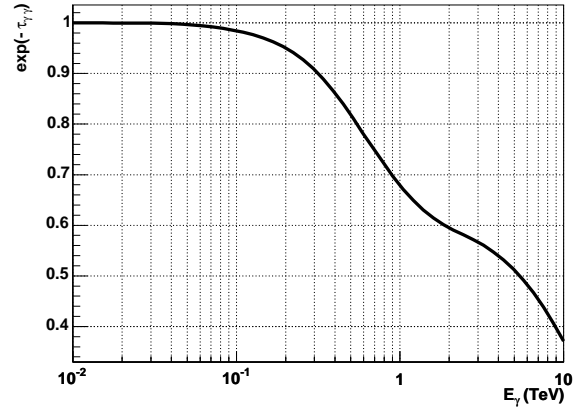}
\caption{\label{tau}
         Attenuation coefficient $exp(-\tau_{\gamma\gamma})$ for Mkn~421 (z=0.030)
         using the EBL spectrum as shown in Fig.~\ref{fig_ebl}.}
\end{minipage}
\end{center}
\end{figure}

\begin{figure*}
\begin{center}
\includegraphics*[width=0.9\textwidth,angle=0,clip]{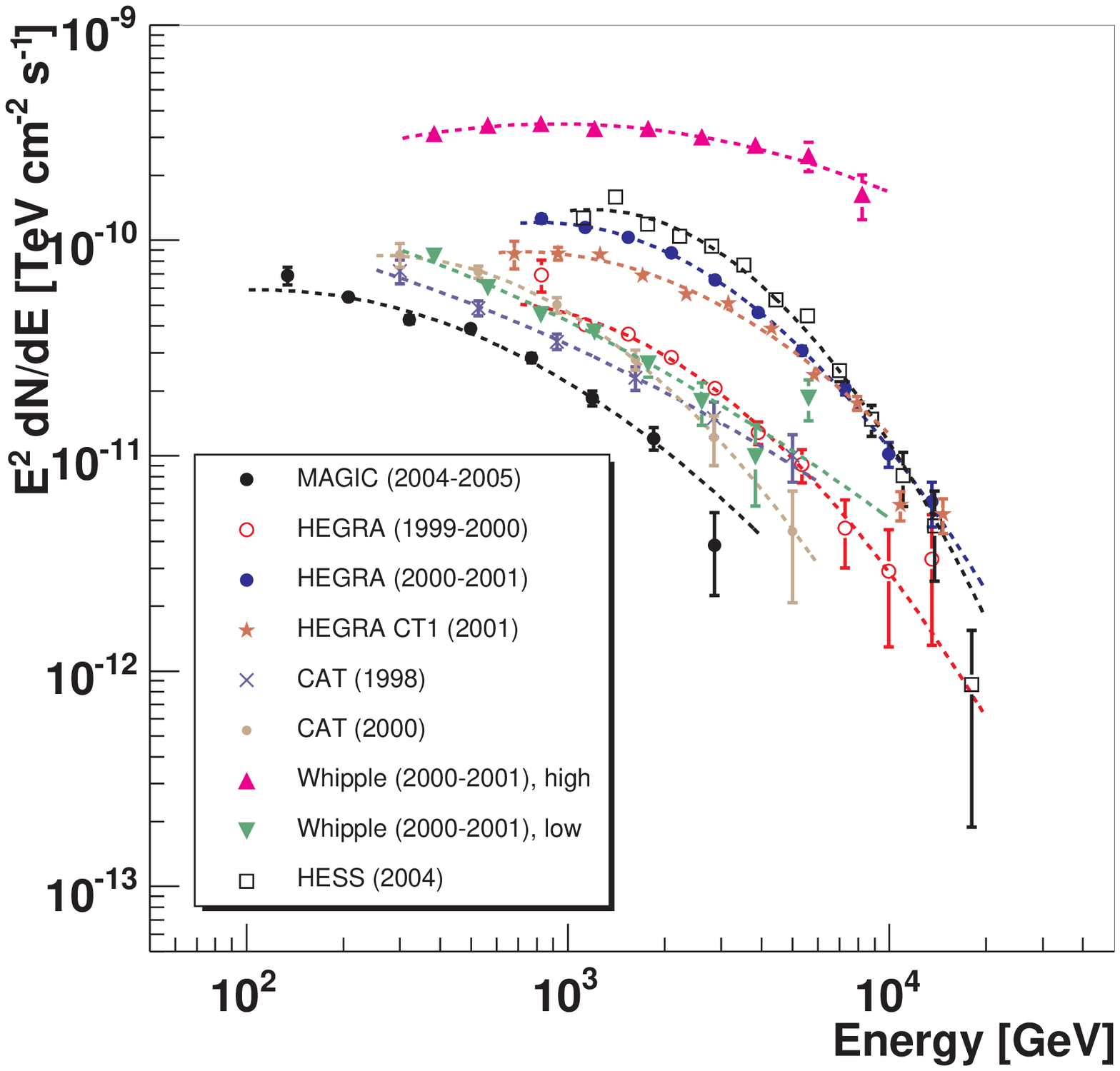}
\caption{\label{fig_allSED}
         Differential energy spectra of Mkn 421 multiplied by $E^2$ 
         in different activity states  from different experiments.
         The spectra are de-absorbed using the EBL model from \citet{primack}, upscaled
         by a factor of 1.5.
         A log-parabolic fit is performed (dashed lines) 
         to determine the peak position in the 
         SED (see Fig.~\ref{fig_ICpeak}). For clarity, only the highest and lowest 
         of the published Whipple results are shown. Note that for the MAGIC spectrum 
         the upper limit at 4.4 TeV is not plotted.}
\end{center}
\end{figure*}

In principle, upper limits on the EBL can also be determined from observed 
$\gamma$-ray spectra from medium to high redshift TeV blazars. Under assumptions 
that the reconstructed TeV blazar spectrum is not too hard and it does not have a pile-up at high
energies, the EBL level can be constrained (see \citet{HauserDwek} for summary 
and \citet{hessebl} for latest results). However, since the measured spectrum of Mkn~421 
is much softer that the one of Mkn~501, which is located at similar redshift, the 
softness seems to be intrinsic. In addition, the data in this paper extend up to 3 TeV only
(historical data of Mkn~421 extend up to 20 TeV), which further weakens possible constraints
from such a nearby source. We therefore do not try to constrain EBL using this Mkn~421
data set. 

Instead, we adopt the recent model of \citet{primack}, scaled up by a factor
1.5 (which is within the model uncertainties), to match lower limits set by the
{\em Spitzer} mission and ISOCAM in the range of 4 to 15~$\mu$m
\citep{spitzer,elbaz,metcalfe}. The resulting EBL spectrum is shown in
Fig.~\ref{fig_ebl} by the black curve.  This EBL spectrum agrees with
alternative models (e.g. \citet{kneiske,pei,blain,steckerNEW}) which are designed to
predict the EBL today.  It is also very close to the upper limits inferred from
arguments on AGN spectra \citep{hessebl}.  Using this EBL spectrum and
state-of-the-art cosmology (flat universe, Hubble constant $H_0$=72 km/s/Mpc,
matter density $\Omega_m$=0.3, dark energy density $\Omega_{\Lambda}$=0.7) we
calculated the optical depth $\tau_{\gamma\gamma}$ for Mkn~421.  Thereby we use
numerical integration of Eq.~2 from \citet{dwek}.  The attenuation
coefficients $exp(-\tau_{\gamma\gamma})$ are shown as the function of energy of
VHE $\gamma$-rays in Fig.~\ref{tau}. 
We note that the attenuation coefficients are very similar to 
those from  \citet{steckerNEW}. 

\subsubsection{The de-absorbed spectrum of Mkn~421}\label{results:intrinsic}

The measured spectrum and the reconstructed de-absorbed (i.e. corrected for the
effect of intergalactic absorption) spectrum are shown in
Fig.~\ref{fig_spectrum}.  For comparison reasons, the Crab Nebula spectrum is
also shown.
The de-absorbed spectrum (shown by filled black
circles) is clearly curved, its probability of being a simple power law is
1.6$\times10^{-8}$.  The de-absorbed spectrum is fitted by a power law 
with an exponential cut-off: 
$ dN/dE = N_{0} (E/0.2\,\mathrm{TeV})^{-\alpha} \exp(-E/E_{\mathrm{cutoff}}) $, $\alpha$
being the photon index, solid line in Fig.~\ref{fig_spectrum}.  The fit
parameters are listed in the inlay of Fig.~\ref{fig_spectrum}.  The power
law with a cut-off describes well the de-absorbed spectrum of Mkn~421, with a photon index
$\alpha = 2.20 \pm 0.08$ and a cut-off energy of $E_{\mathrm{cutoff}} = (1.44 \pm 0.28)$\,TeV.  
Taking into account the systematic uncetrainty of 18\% on the absolute energy scale
of our measurement and in addition a guessed 25\% uncertainty on the EBL level,
we find that neither the photon index nor the cut-off energy 
substantially change (See Table~\ref{tab:systcutoff}). The fitted 
photon index was found to be between 2.12 and 2.24, whereas the cut-off energy 
was found to be between 1.1 and 1.6~TeV.  
From this study we conclude that the curvature of the spectrum is source inherent: 
either at the measured flux level this cosmic accelerator is close to its energy limit, 
or there exists a source--intrinsic absorption.


\section{Discussion}\label{discus}

\subsection{Comparison with previous observations of Mkn~421}

\begin{figure}
\begin{center}
\begin{minipage}{0.45\textwidth}
\includegraphics*[width=1\textwidth,angle=0,clip]{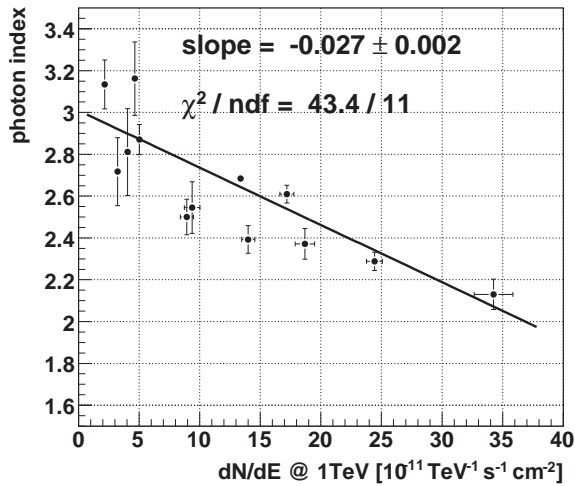}
\caption{\label{fig_hardness}
         Relation between the flux state at 1\,TeV (determined from a power law approximation
         of the spectra between 700 GeV and 4 TeV) and the fitted photon index of 
         published data as in Fig.~\ref{fig_allSED}.
         A correlation between flux and hardness can be clearly seen. }
\end{minipage}
\end{center}
\end{figure}

\begin{figure}
\begin{center}
\begin{minipage}{0.45\textwidth}
\includegraphics*[width=1\textwidth,angle=0,clip]{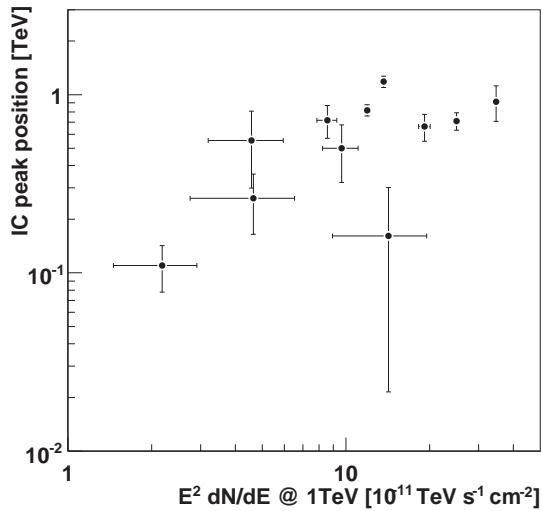}
\caption{\label{fig_ICpeak}
         Relation between the fitted peak position in the SED and the energy density 
         at 1 TeV for published data as in Fig.~\ref{fig_allSED}. 
         A clear trend can be observed for the peak position to
         shift towards higher energies with increased source intensity.}
\end{minipage}
\end{center}
\end{figure}

In Fig.~\ref{fig_allSED} we show the (de-absorbed) energy density spectrum 
in context with several previously published high statistics observations of Mkn 421.
For a compilation of the VHE measurements of Mkn~421 we used historical
data from CAT \citep{piron421}, HEGRA \citep{hegra421,ct1_421}, HESS \citep{hess421}, and  
Whipple \citep{whipple421}\footnote{For the Whipple measurements, only spectra in the highest 
and in the lowest flux state are shown in order not to clutter Fig.~\ref{fig_allSED}}.  
All measured spectra are de-absorbed using the EBL model
as described in Subsection~\ref{results:ebl}.
The activity of the source during MAGIC observations clearly was at the lower end,
and our results extend to energies lower than those previously observed, thus
being complementary both in source intensity and energy range. All results
seem consistent with each other, and all show significant deviations from a
simple power law, which can not be explained by attenuation effects (the results are 
robust with respect to the EBL model within a factor $\pm$25\%). They are, 
therefore, likely to be source--intrinsic. 

From the compilation of the de-absorbed Mkn~421 spectra, it is evident that with 
an increasing flux state the spectrum becomes harder. In order to verify this, we fitted the
spectra by a simple power law ($\mbox{dN}/\mbox{dE} \propto \mbox{E}^{-\alpha}
$) in the overlapping energy region between 700 GeV and 4 TeV.  The resulting
photon indices $\alpha$ as function of the fitted flux at 1 TeV are shown in
Fig.~\ref{fig_hardness}.  Evidently, with increasing flux the spectra harden.
Similar results were obtained using Whipple data \citep{whipple421}, HEGRA data 
\citep{hegra421,kranich}, and CAT data \citep{cat421}.

The curvatures observed are indicative of a maximum in energy density, 
and are usually interpreted
as due to inverse-Compton (IC) scattering. The peak position appears to be dependent on the 
source flux intensity. We have, therefore, 
performed a log-parabolic fit for all available data. The log-parabola has the following 
parametrization:
  $\log10(\nu F_{\nu}) = \mbox{A} + \mbox{B} \left(\log10(\mbox{E}) - \log10(\mbox{E}_p) \right) $, 
with $\nu F_{\nu} = \mbox{E}^2\, \mbox{dN}/\mbox{dE}$ and $\mbox{E}_p$ being the energy of the peak position.
The best log-parabolic fits are shown in Fig.~\ref{fig_allSED} by the dashed lines.  
In Fig.~\ref{fig_ICpeak} we compare the resulting peak positions for 
the different experiments as a function of their (fitted) energy density at 1~TeV. Evidently,
with increasing flux the peak shifts to higher energy values.
Future observations at higher intensities extending to lower energies will
have to corroborate these results. Such observations are 
part of the future MAGIC physics progamme.

\subsection{A short comment on the light curves}

In the observation period between November 2004 and April 2005 we observed night-to-night 
flux variations up to a factor of 2 and a maximum flux change in the entire set of a factor 4.
No short-term flux variations well below 1 hour, as observed during high flaring activity
in previous experiments \citep{gaidos,hegra421}, were seen, although the sensitivity of MAGIC
would allow to detect fast flares in the given flux range. Two equally likely explanations
are that either we deal with large fluctuations resulting in the absence of any fast flare
during the observation period, or fast flaring is a feature that occurs only when
the source is very active. This calls for further high statistics and high sensitivity
studies when the source is in its low flux state.

\subsection{Correlation studies}

\begin{figure}
\centering
\includegraphics*[width=0.46\textwidth,angle=0,clip]{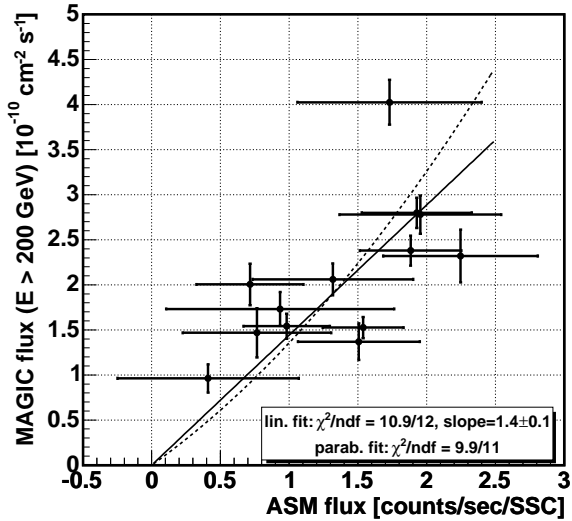}
\caption{\label{fig_correl_xrays}
	Correlation between MAGIC integral flux measurements above 200~GeV 
    and RXTE/ASM counts for 13 nights.}
\end{figure}

The correlation between the $\gamma$-ray flux measured by MAGIC and the
X-ray flux measured by RXTE/ASM is shown in Fig.~\ref{fig_correl_xrays}. 
For the MAGIC flux 
we take the nightly average above 200~GeV (see also Fig.~\ref{fig_lightcurve}).
For the X-ray data, we calculate the average of those RXTE/ASM pointings (dwells) which
were taken simultaneously with MAGIC data, allowing $\pm$1~hour with
respect to the MAGIC data, to increase X-ray statistics.
Fig.~\ref{fig_correl_xrays} shows a clear correlation between X-ray and
$\gamma$-ray data.  The linear fit (solid line), forced to go through (0,0),
has a slope of 
$1.4\,\pm\,0.1\, [\frac{10^{-10}}{\mbox{\small cm}^2}/\frac{\mbox{\small counts}}{\mbox{\small SSC}}]$, 
and has a $\chi^2$ probability of 54\%. 
The parabolic fit (dashed line) which is
also forced to go through (0,0) has the same $\chi^2$ probability of 54\%. 
The correlation coefficient $r=0.64^{+0.15}_{-0.22}$ 
(errors correspond to 1\,$\sigma$ level) is different from zero by 2.4 standard deviations
(taking into account the non-linearity of errors).

\begin{figure}
\centering
\includegraphics*[width=0.46\textwidth,angle=0,clip]{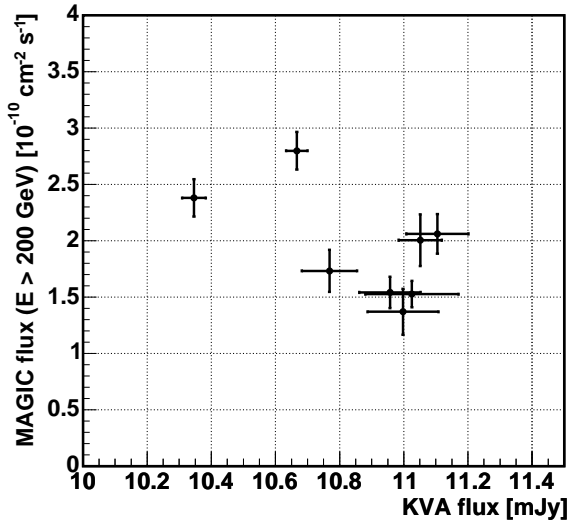}
\caption{\label{fig_correl_optic} 
	Correlation between MAGIC integral flux above 200~GeV 
    and optical flux measured by the KVA telescope for 8 nights.}
\end{figure}

In Fig.~\ref{fig_correl_optic} the MAGIC $\gamma$-ray flux above 200~GeV is
shown together with simultaneous KVA optical data.  The
latter have been averaged over the MAGIC integration time. 
One can see a possible $\gamma$-ray/optical anticorrelation  
during the 8 nights of simultaneous observations, 
however, the correlation coefficient $r=-0.59^{+0.36}_{-0.22}$ 
is compatible with zero within 1.5 standard deviations.


\subsection{Comparison with models}
 
\begin{figure*}
\begin{center}
\includegraphics*[width=0.45\textwidth,angle=0,clip]{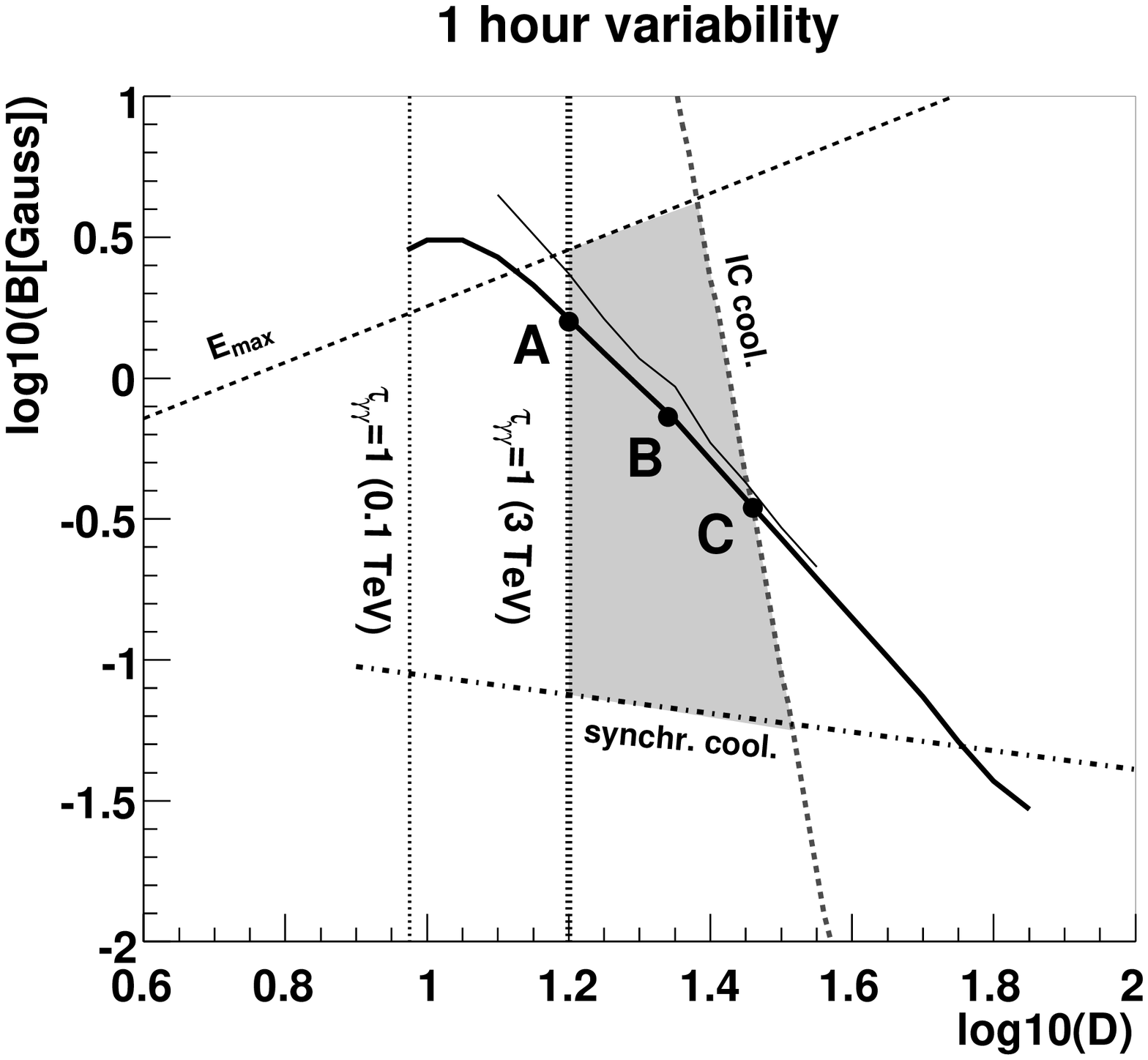}
\includegraphics*[width=0.45\textwidth,angle=0,clip]{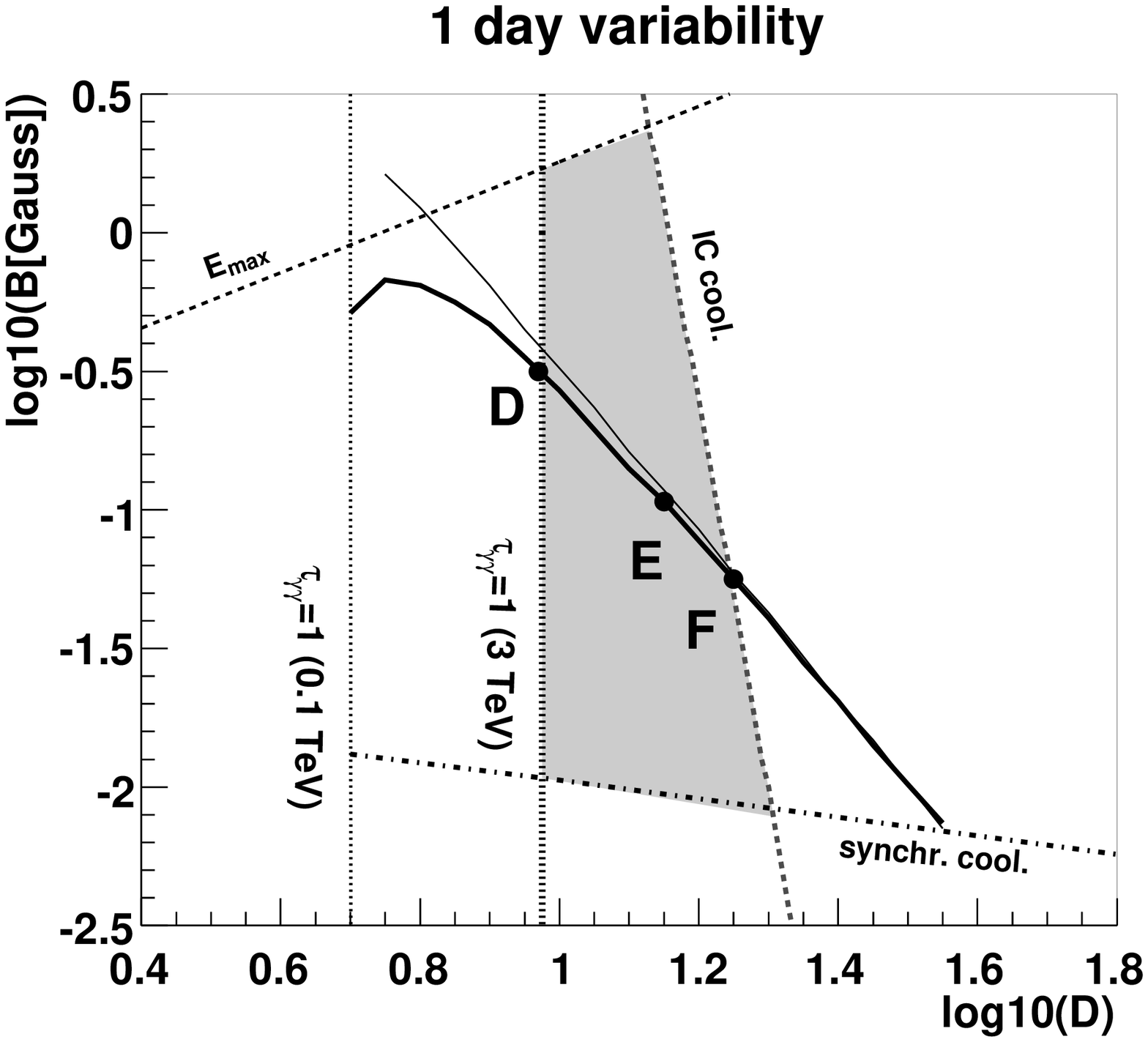}
\hfill
\includegraphics*[width=0.46\textwidth,angle=0,clip]{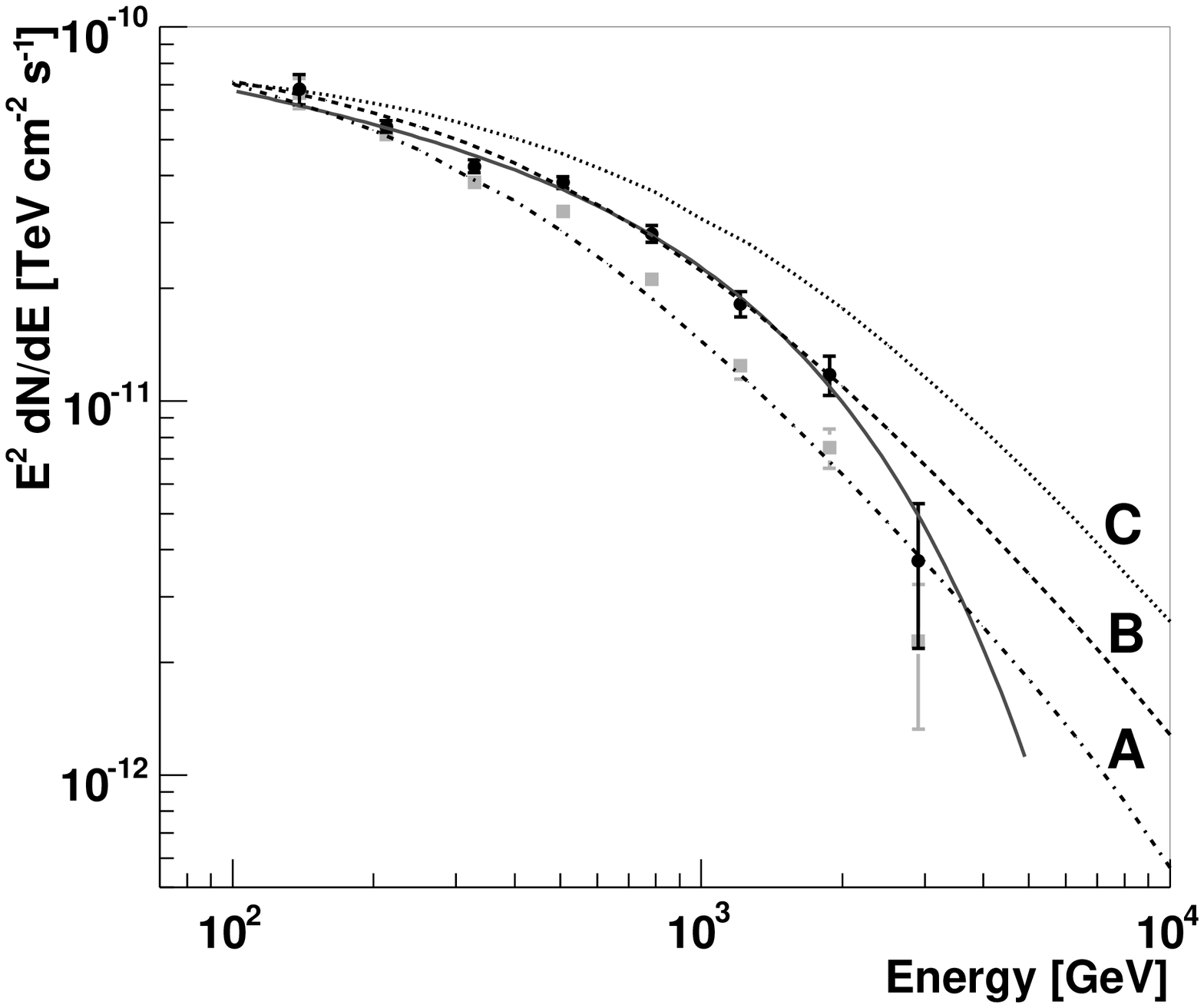}
\includegraphics*[width=0.46\textwidth,angle=0,clip]{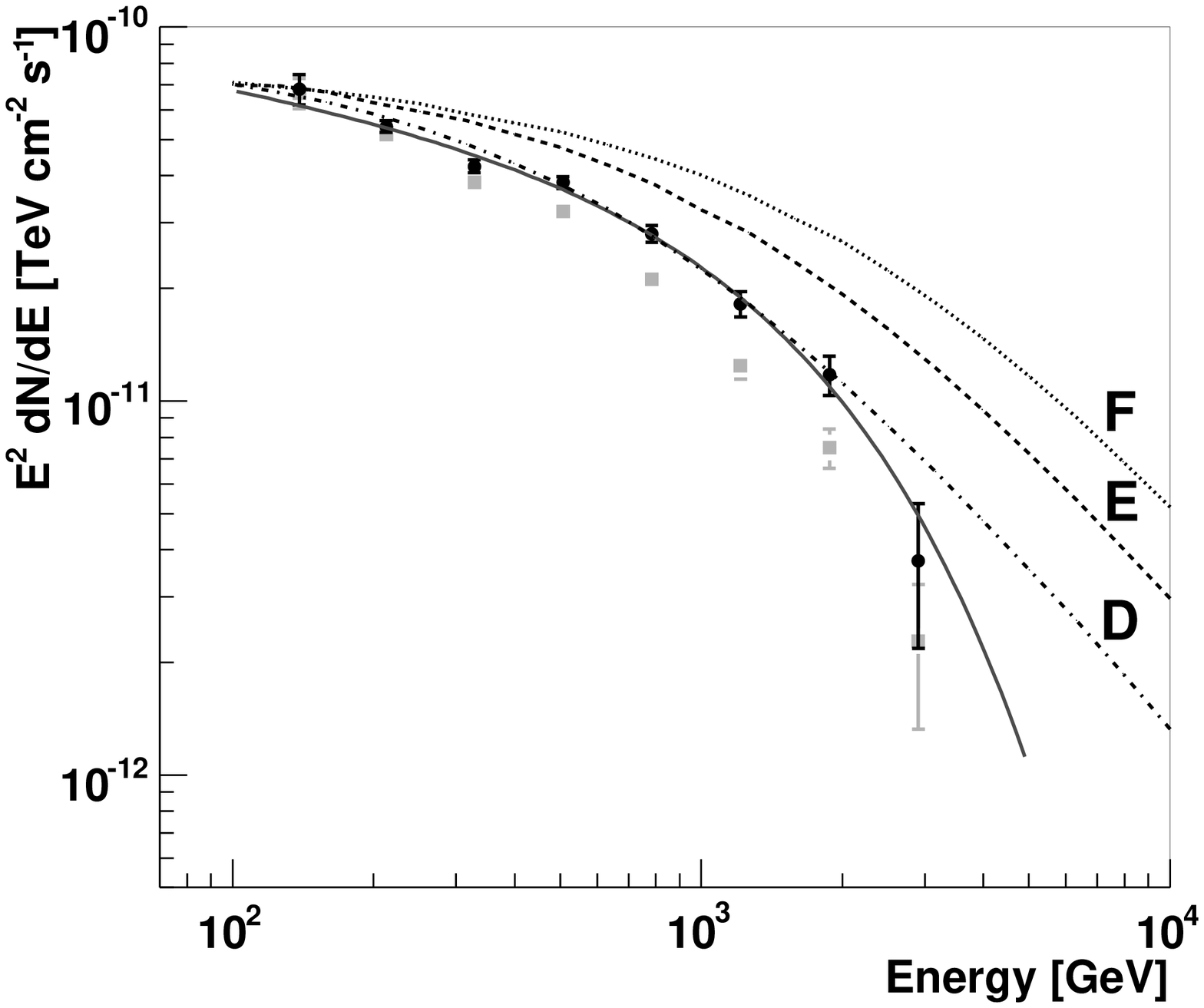}
\caption{\label{ssc_constraints} 
Constraints on the parameter space (Doppler factor, $D$, versus  magnetic field
strength, $B$) for the emission region in the jet of Mkn~421 based on the one-zone
homogeneous SSC model.  From the ratio of the $\gamma$-ray  to X-ray power
during the MAGIC observations of Mkn~421 (measured at the peaks in the
synchrotron and IC spectra) the allowed values are limited to the thick curves
(assumed IC peak at 100~GeV) or to the thin curves (assumed IC peak at 10~GeV).
The left figures correspond to a 1~hour variability, whereas the right figures correspond
to a 1~day variability. 
The physical conditions in the blob are limited by the electron cooling on the synchrotron and IC
processes, optical depth for $\gamma$-rays, and maximum energy of electrons as discussed in the text.
These constraints are shown by the dashed and dot-dashed lines.
The allowed region is limited by these lines and 
is marked by the grey shaded area. The
$\gamma$-ray spectra are calculated for the values of the Doppler factor and
magnetic field strength marked by A, B, and C (for 1~hr variability), and D, E,
and F (for 1~day variability). They are compared with the de-absorbed MAGIC spectrum 
(shown as full black circles) in the bottom figures. The fit by a power law with an exponential 
cut-off to the de-absorbed spectrum (as in Fig~\ref{fig_spectrum}) is shown by the black solid line, 
whereas the measured spectrum is shown by the grey full squares.}
\end{center}
\end{figure*}

\begin{figure}[t]
\begin{center}
\begin{minipage}{0.45\textwidth}
\includegraphics*[width=1.0\textwidth,angle=0,clip]{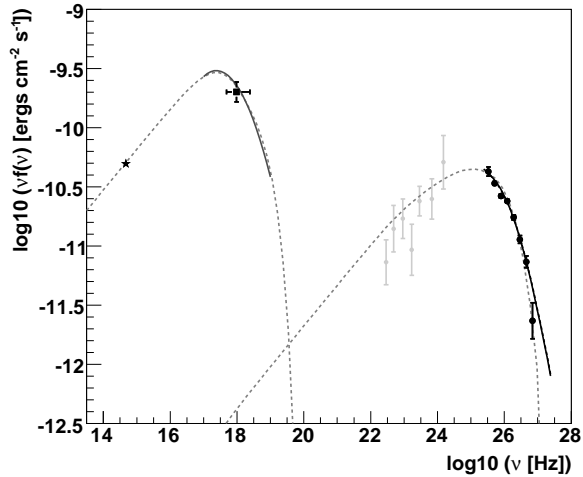}
\caption{\label{figure_sedfit}
         The overall SED of Mkn~421 from optical wavelengths through VHE $\gamma$-rays. 
         Large symbols represent averaged data described in this paper:
         optical data from KVA (star), 
         X-rays from RXTE/ASM (full square), 
         de-absorbed $\gamma$-rays from MAGIC (full points).
         The grey full squares are archival EGRET measurements \citep{3egret}.
         The grey curve in the X-rays corresponds to the
         log-parabolic fit taken from \citet{massaro}
         using {\em Beppo}SAX \citep{bepposax} data of Mkn~421 taken on
         21 April 1998. The two black curves through the $\gamma$-ray spectrum 
         (almost indistinguishable) correspond to the SSC model parameter sets
         B and D (see text and Fig.~\ref{ssc_constraints} for details).
         The grey dashed line denotes a fit by the SSC model as in \citet{krawmodel}, 
         see text for details.}
\end{minipage}
\end{center}
\end{figure}

Given the temporal correlation between X-ray and $\gamma$-ray fluxes, it is
reasonable to infer that the VHE $\gamma$-ray radiation is dominated by
emission resulting from IC upscattering of the 
synchrotron X-ray photons by their parent population of relativistic electrons.
Such correlation can be modelled with a homogeneous synchrotron-self-Compton (SSC) model. 
Based on this model it is possible to constrain the parameter space of the
emission region and estimate its basic parameters, the Doppler factor, $D$, and 
the rest-frame magnetic field, $B$, of the emitting plasma in the relativistic jet.
To this end we follow the procedure first devised by \citet{bednarek97}
for the Mkn~421 flare of 16 May 1994, subsequently improved by, e.g., 
\citet{tavecchio,bednarek99,kataoka,katarzynski}.
Application of this method requires precise simultaneous multiwavelength information.
Since a synchrotron (X-ray) spectrum simultaneous with the MAGIC observations is not available,
we have to resort to previous X-ray observations arguing that similar TeV $\gamma$-ray
states (IC emission) should correspond to similar X-ray states (synchrotron emission).
In fact, similar $\gamma$-ray spectra of Mkn~421 have already been observed 
several times -- including the HEGRA observations in April 1998 \citep{ahar1999} 
for which simultaneous {\it Beppo}SAX observations are available \citep{fossati,massaro}. 
Here we use the X-ray spectra and parameterization, reported by \citet{massaro} for 
21 April 1998. It is also noticeable that the X-ray flux level between the
simultaneous RXTE/ASM data and the BeppoSAX data used here is very similar
(see Fig.~\ref{figure_sedfit}).

The low flux state MAGIC $\gamma$-ray spectrum, reported here for energies at
$\sim$100~GeV, warrants a better investigation of the crucial energy range 
where the IC peak is expected to occur,
than in previous data sets. Following \citet{bednarek97,bednarek99}
we then constrain the allowed parameters of the emission region ($D$ and $B$) from the ratio 
of the $\gamma$-ray power to the X-ray power, measured at their respective peak emission 
(see thick curves in the upper panels of Fig.~\ref{ssc_constraints}). 
The radiation field density and the electron spectrum, cospatial in the blob,
were derived as a function of $D$ and $B$ for a blob radius assumed 
equal to the light travel corresponding to the shortest reported 
variability time scale (for observational arguments see \citet{takahashi}).  
We further constrain the allowed parameter space by arguing that the synchrotron and IC cooling
time scales should be shorter than the observed variability time scale. These conditions are
fulfilled above the dot-dashed lines (for synchrotron cooling) and on the left of
the grey dashed line (for the IC cooling) for the 1~hr 
(upper left panel of Fig.~\ref{ssc_constraints}) and 1~day 
(upper right panel of Fig.~\ref{ssc_constraints}) variability time scales. 
The condition that the blob has to be transparent to the VHE $\gamma$-rays leads
to a further lower bound on $D$ by requiring that the optical depth by pair production
has to be lower than unity. The corresponding limits for photon energies of 100~GeV
and 3~TeV (which define the energy range of MAGIC measurement) are shown in the 
upper panels of Fig.~\ref{ssc_constraints} as
thin and thick dotted lines, respectively.
One last condition arises from comparing the maximum energy of electrons,
determined by the maximum energy of synchrotron photons $\sim$40~keV, with the maximum 
energy of the detected photons $\sim$3~TeV (see dashed line in the upper panels of
Fig.~\ref{ssc_constraints}). These limiting conditions build an allowed region in the 
$D$-$B$ plane as marked by the grey shaded area.
The allowed parameters of the emission region correspond to the part of the
thick full curve inside the region limited by all these lines (see
Fig.~\ref{ssc_constraints}). In order to determine the values of $D$ and $B$ more
precisely, we now calculate the $\gamma$-ray spectra for the points A, B, and C
for 1~hr variability, and the points D, E, and F for 1~day variability, and
compare the predicted spectrum with the actual de-absorbed spectrum.  From the
lower panels of Fig.~\ref{ssc_constraints} it is clear that the best
description is provided by the blob with Doppler factor $D\sim$22 and magnetic
field $B\sim$0.7~G (the point B) for 1~hr variability, and $D\sim$9 and
$B\sim$0.3~G (the point D) for 1 day variability.  In order to assess how this
result is sensitive on the correct energy localization of the peak in the
$\gamma$-ray spectrum (which is in fact only limited by the lower energy end of
the MAGIC spectrum), we show the allowed parameter space for the $\gamma$-ray
peak at 10~GeV (see thin full curves in Fig.~\ref{ssc_constraints}). The
constraints for the peak at 10~GeV and 100~GeV are almost the same for the
parts of the curves inside the allowed region.  It is interesting that the
emission parameters, as  estimated here for the low flux state of Mkn~421
(for the 1~day variability time scale), are not very different from those
estimated by \citet{bednarek97} for the flaring state.  This suggest that the
flaring state may not be related to the significant change of the blob's
Doppler factor and magnetic field strength. 

In Fig.~\ref{figure_sedfit} we show
the broadband SED of Mkn~421.  Large symbols represent averaged data described
in this paper: optical data from KVA (star), X-rays from RXTE/ASM (full
square), $\gamma$-rays from MAGIC (full points). The grey curve in the X-rays
corresponds to a log-parabolic fit performed by \citet{massaro} on {\em
Beppo}SAX data of Mkn~421 taken on 21 April 1998. The two black curves through
the $\gamma$-ray spectrum are almost indistinguishable and correspond to the
best SSC model parameters for 1-hr and 1-day variability time scales (points B
and D respectively, calculated according to Eq.~(13) in ~\cite{bednarek99} 
who apply the Klein-Nishina cross-section as in~2.48 of~\cite{blumenthal70}).

\begin{table}[h]
\begin{center}
\caption{\label{tab:ssc} SSC model parameters for Mkn~421. The corresponding photon spectrum
is shown in Fig.~\ref{figure_sedfit}.}
\begin{tabular}{ l c }
\hline
 \multicolumn{2}{c}{spherical blob with:} \\
 Doppler factor   & 15 \\
 magnetic field  & 0.20 Gauss \\
 radius of emitting region & $1.6\times10^{16}$~cm \\
\hline
 \multicolumn{2}{c}{injected electron spectrum:} \\
 electron energy density & 0.06 erg/cm$^3$ \\
 5 $<$ log10(E[eV]) $<$ 10.9 & index 2.31 \\
 10.9 $<$ log10(E[eV]) $<$ 11.6 & index 3.88 \\
\hline
\end{tabular}
\end{center}
\end{table}

In addition, we apply the SSC code provided by \citet{krawmodel} to our dataset.
The fitted overall SED is shown by the grey dashed line in Fig.~\ref{figure_sedfit}, 
and the model parameters are listed in Table~\ref{tab:ssc}. 
For the fit, we used the simultaneous KVA, ASM and MAGIC data, as well as the 
archival {\it Beppo}SAX observations from 21 April 1998 \citep{massaro}.
In contrast to the parameters adopted in \citet{krawczynski}, we used a smaller Doppler factor 
(15 instead of 50), resulting in a somewhat larger emitting region (1.6$\times10^{16}$~cm 
instead of 2.7$\times10^{15}$~cm), and a higher particle density (0.06~erg/cm$^3$ instead
of 0.01~erg/cm$^3$).
We note that the fitted values of magnetic field and Doppler factor are within the
allowed range as defined above. 
Remarkably, the archival EGRET data \citep{3egret} matches the fit almost perfectly,
 suggesting an IC peak around 100~GeV.

\section{Concluding Remarks}

Mkn~421 was observed with the MAGIC telescope during several months in 2004 and
2005. Briefly, we have presented the following:
\begin{itemize}
 \item  first high-sensitivity observation down to $\approx$ 100\,GeV;
 \item  first observation of an IC peak at low flux;
 \item  absence of short flares below 1 hour duration despite sufficient
        sensitivity;
 \item  flux variation up to a factor 2 between consecutive nights 
        and up to a factor 4 in the entire observation period;
 \item  confirmation of a source--inherent effect resulting in a curved spectrum after 
        de-absorption (for reasonable assumptions concerning the EBL)
        in case of low flux intensity;
 \item   a strong correlation between spectral hardness 
         (photon index between 700\,GeV and 4\,TeV) and flux intensity,
        obtained by comparison of the de-absorbed energy spectra of various experiments covering
        different flux levels;
 \item  a clear trend for the peak position to shift towards higher energies with increased source intensity,
        obtained by the same comparison;
  \item confirmation of a significant correlation between X-ray and VHE $\gamma$-ray
        intensity during a state of low to medium intensity;
  \item a hint that different flaring states result from differences in electron
        populations (electron spectrum) rather than from significant change of the blob's
        Doppler factor and magnetic field strength.

\end{itemize}

We add the following conclusions.
The flux state was found to be comparatively low, 
ranging in intensity between 0.5 and 2 Crab units when
integrated above 200\,GeV.  
While clear night to night variations were found, the intra-night light curve, binned in 
10-minute time intervals, does not show significant variations, although several nights
are only marginally compatible with a constant flux. They do not show a discernible structure,
though, and seem not associated to an overall flux different from that of
perfectly quiescent nights.
We note that MAGIC is sensitive enough to
detect variabilities on the 10-minute time scale at such a low flux level.
A clear correlation ($r=0.64^{+0.15}_{-0.22}$) between X-rays and $\gamma$-rays was found, while  
no correlation was seen between optical and $\gamma$-rays.  This supports a leptonic
origin of the $\gamma$-rays from Mkn~421. 
The energy spectrum resulting from the combined MAGIC data, corrected for
the extragalactic absorption, suggests the presence of an IC peak at about 100~GeV.
The spectrum is clearly curved at energies above 1~TeV, and can be fitted by a power-law
with an exponential cut-off.
The overall SED observed in the observed flux state can be well described by 
a homogeneous SSC model provided that the emission region moves with a Doppler factor 
$\sim$9 and its magnetic field strength is $\sim$0.3~G for a 1-day 
variability time scale. Surprisingly, these parameters do not differ
substantially from those estimated for the emission region of Mkn~421 during a strong flare
\citep{bednarek97}.
The fit with an alternative SSC code of \citet{krawczynski} lead to similar 
Doppler factor and magnetic field values.

\acknowledgements
We would like to thank the IAC for the excellent working conditions at the
Observatory de los Muchachos in La Palma. The support of the German BMBF and
MPG, the Italian INFN and the Spanish CICYT is gratefully acknowledged. This
work was also supported by ETH Research Grant TH~34/04~3 and the Polish MNiI
Grant 1P03D01028. We also thank Dieter Horns and Frank Krennrich
for providing us with HEGRA, H.E.S.S., and Whipple data.

\end{document}